\documentclass[aps,prl,twocolumn,superscriptaddress]{revtex4}

\usepackage{graphicx}

\begin{document}


\title{Carrier-mediated antiferromagnetic interlayer exchange coupling in diluted magnetic semiconductor multilayers Ga$_{1-x}$Mn$_x$As/GaAs:Be}


\author{J.-H. Chung}
\author{S. J. Chung}
\author{Sanghoon Lee}\email{slee3@korea.ac.kr}
\affiliation{Department of Physics, Korea University, Seoul 136-713, Korea}
\author{B. J. Kirby}
\author{J. A. Borchers}
\affiliation{National Institute of Standards and Technology, Gaithersburg, MD 20899, U.S.A.}
\author{Y. J. Cho}
\author{X. Liu}
\author{J. K. Furdyna}
\affiliation{Department of Physics, University of Notre Dame, Notre Dame, IN 46556, U.S.A.}


\date{\today}

\begin{abstract}

We use neutron reflectometry to investigate the interlayer
exchange coupling between Ga$_{0.97}$Mn$_{0.03}$As ferromagnetic
semiconductor layers separated by non-magnetic Be-doped GaAs
spacers. Polarized neutron reflectivity measured below the Curie
temperature of Ga$_{0.97}$Mn$_{0.03}$As reveals a characteristic splitting at the wave vector corresponding to twice the multilayer period, indicating that the
coupling between the ferromagnetic layers are antiferromagnetic (AFM).
When the applied field is increased to above the saturation field, this AFM coupling is suppressed. This behavior is not observed when the spacers are undoped,
suggesting that the observed AFM coupling is mediated by
charge carriers introduced via Be doping. The behavior of
magnetization of the multilayers measured by DC magnetometry is
consistent with the neutron reflectometry results.

\end{abstract}


\maketitle


The exploration of systems that combine electronic and spin
degrees of freedom is the subject of major interest in recent
semiconductor electronics research. Spin dependent transport has
already demonstrated its technological impact in the form of
metallic ferromagnetic multilayers, where giant resistance changes
are observed under external magnetic
field.\cite{baibich88,binasch89} A prerequisite for such large
magnetoresistance is the presence of stable antiferromagnetic
(AFM) coupling between ferromagnetic (FM) layers, which can be
overcome by the application of an applied field.\cite{grunberg86}
Spin-dependent scattering of charge carriers that changes greatly
depending on interlayer exchange coupling (IEC) is the origin of
the observed magnetoresistance. Such AFM IEC has been observed in
various metallic\cite{parkin90,parkin91,bloemen94,borchers99} and
semiconductor\cite{rhyne98,kepa01el,kepa03pb,kepa03prb}
multilayers.

Diluted magnetic semiconductors (DMSs) in which ferromagnetism is
induced via spin-charge doping into III-V semiconductors have
been widely studied with an eye on combining spintronics with
well-established semiconductor technology. One of the most
intensively studied DMS systems is
Ga$_{1-x}$Mn$_x$As,\cite{jungwirth06} where the substitutional
doping by Mn$'_\mathrm{Ga}$ results in ferromagnetism mediated by
holes.\cite{beschoten99,blinowski03} Realization of reversible
switching between AFM and FM spin states in DMS multilayers may
greatly enhance the magnetoresistance in these systems. So far, however, AFM IEC in GaMnAs-based multilayers has never been reported, and only the FM IEC is explicitly observed.\cite{munekata92,ohno92,ohno96,hayashi97,kepa01prb,kirby07}. In principle, AFM IEC is expected to emerge when carrier density is enhanced in the non-mangetic spacers.\cite{jungwirth99,sankowski05,giddings06}
Several recent experimental studies indeed suggested possible signatures of weak or partial AFM IEC via indirect modulation-doping.\cite{ge07,kirby08} Therefore, it is suspected that carrier doping directly into the spacers has high promise for achieving robust AFM IEC.

In this work, we have used polarized neutron reflectometry to obtain definitive evidence of AFM IEC in a
DMS-based multilayer structure Ga$_{0.97}$Mn$_{0.03}$As/GaAs:Be in which the nonmagnetic spacers are doped by Be. Importantly, FM alignment was only observed in the case of a sample with undoped spacers, indicating that the AFM IEC was
mediated by charge carriers in the nonmagnetic spacers introduced
via Be doping. We show additionally that DC magnetization
measurements are also consistent with the IEC described above.

The DMS multilayers used in this study consist of ten
Ga$_{0.97}$Mn$_{0.03}$As layers separated by GaAs spacers,
deposited on GaAs (001) substrates by molecular beam epitaxy. Two
nearly identical samples were fabricated, one with Be-doped GaAs spacers,
the other with undoped spacers. The Be concentration in the
spacers is estimated as 1.2$\times10^{20}$/cm$^3$ from Hall
measurements carried out on a reference sample. A capping layer of
undoped GaAs was deposited on top of both multilayers. From
neutron reflectivity data (discussed later), the thicknesses of
DMS, spacer, and capping layers are estimated to be
6.95 nm, 3.47 nm, and 3.47 nm (2.90 nm), respectively, for the Be-doped (undoped) sample.

\begin{figure}
\includegraphics[width=3.0in]{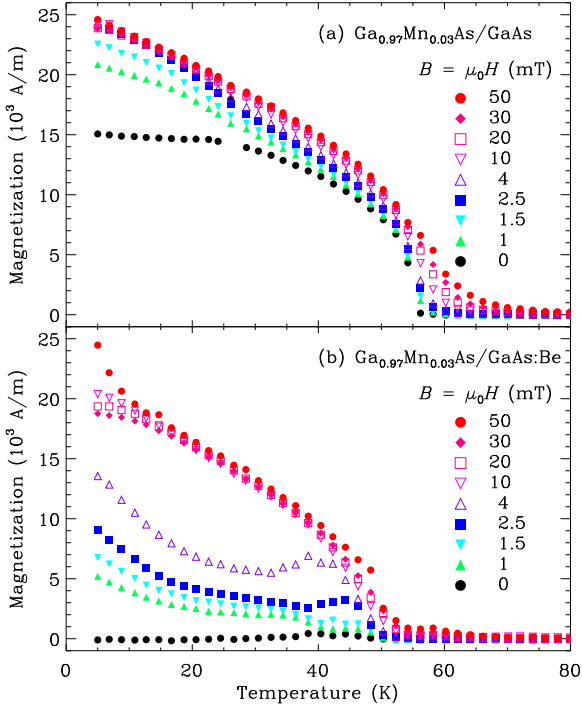}%
\caption{\label{magnetization} Temperature dependence of
magnetization of Ga$_{0.97}$Mn$_{0.03}$As/GaAs(a) and Ga$_{0.97}$Mn$_{0.03}$As/GaAs:Be(b) multilayers. The data were collected while cooling, with the magnetic field applied along the [100]
direction.}
\end{figure}

We begin by presenting the temperature dependences of the
magnetization of the multilayers measured using a SQUID vibrating
sample magnetometer while cooling. Figure \ref{magnetization}
shows the magnetization curves in a series of DC fields applied
along the [100] direction, which is approximately parallel to the
magnetic easy axis at the lowest temperature. In the sample with undoped
spacers, the magnetization increases below $T \approx$ 60 K
following the typical behavior of ferromagnetic
Ga$_{1-x}$Mn$_x$As. A small kink is observed at $T \approx$ 28 K,
suggesting the development of a biaxial cubic anisotropy.\cite{shin07} In sharp contrast, in the sample with Be-doped spacers the temperature behavior of magnetization measured in low fields is very different. For instance, the magnetization
measured at 4 mT rises around 50 K, then drops as the temperature
is lowered to below 40 K, followed by another upturn. At zero
field, the net magnetization is almost completely suppressed,
showing only a very weak signal below $T_C$. Such a large decrease
in net magnetization indicates significant changes in exchange
coupling due to Be doping, strongly suggesting that IEC between
ferromagnetic layers is antiferromagnetic. In comparison, the
magnetization of the multilayers measured at fields higher than 10
mT shows normal FM behavior, suggesting that in these fields IEC is
ferromagnetic.

To confirm the presence of the suspected AFM IEC, we performed
unpolarized and polarized neutron reflectivity measurements. The
experiments were done using the NG-1 Polarized Beam Reflectometer at
the NIST Center for Neutron Research. We note that during the
polarized reflectivity measurements it was necessary to apply a
magnetic field higher than 1 mT in order to maintain high neutron
spin polarization ($>$ 93 \%). The samples were oriented with the
[110] axis parallel to the polarization of a monochromatic neutron
beam ($\lambda$ = 4.75 \AA). Four polarized magnetic scattering
intensities were separately measured, from which fully reduced
reflectivity curves were obtained.\cite{majkrzak91} Two non-spin
flip (NSF) structure factors, ($++$) and ($--$), are written in the Born Approximation as
$F^{(\pm\pm)}(Q)=\sum^N_{j=1}(b_j\mp p_j\cos\phi_j)e^{iQu_j}$,
where $b_j$ and $p_j$ are the nuclear and the magnetic scattering
lengths, respectively, $\phi_j$ is the angle between magnetization
vector and applied field, and $u_j$ is the position of the $j$-th
atomic plane. The magnetization components parallel to the applied
field can then be obtained from the splittings between the two NSF
intensities. The spin flip (SF) structure factors, ($+-$) and
($-+$), on the other hand, are given by
$F^{(\pm\mp)}(Q)=\pm\sum^N_{j=1}p_j\sin\phi_je^{iQu_j}$, and
involve magnetization components perpendicular to the applied
field. Below we show only the NSF reflection intensities, because
the SF intensities for the DMS layers are much weaker (by several orders of
magnitude).

\begin{figure}
\includegraphics[width=3.2in]{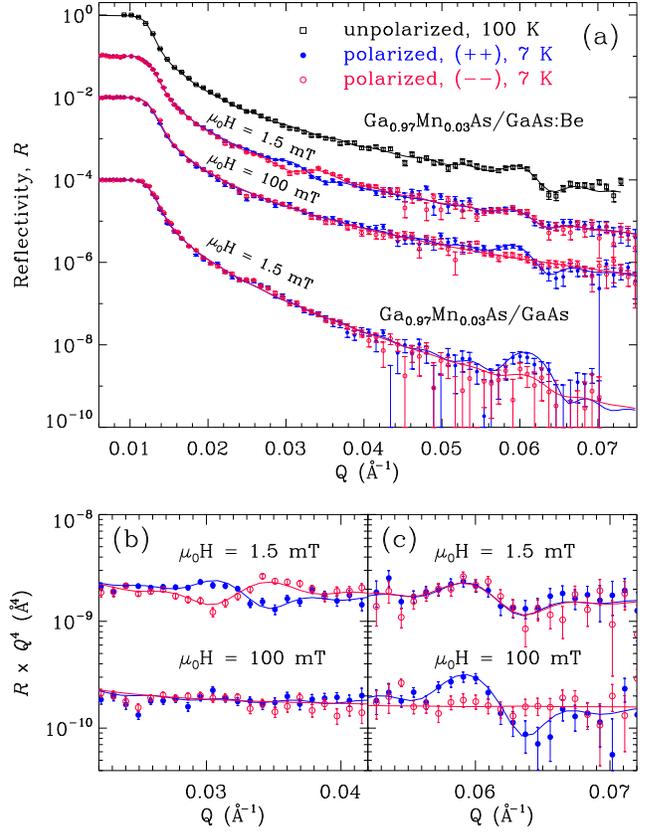}%
\caption{\label{reflectivity} (a) Unpolarized and polarized
neutron reflectivities of the Be-doped and the undoped DMS-based
multilayers with the field applied along the [110] direction. The
curves are shifted vertically for clarity. The unpolarized and the
polarized data were measured at 100 K and 7 K, respectively. The
solid lines are fits to the data using the models described in the
text. (b) and (c) show the reflectivity of the Be-doped multilayer
multiplied by Q$^4$, emphasizing the AFM(b) and the FM(c)
splittings, respectively.}
\end{figure}

The unpolarized neutron reflectivity measured on the Be-doped
sample at 100 K (i.e., when the GaMnAs layers are in the
paramagnetic phase), plotted in the uppermost part of Figure
\ref{reflectivity}(a), shows a structural Bragg peak located at
$\approx$ 0.062 \AA$^{-1}$, which corresponds to the multilayer
periodicity of $\approx$ 100 \AA. The S-shaped profile of the
Bragg peak (instead of a simple peak) is attributed to the presence of a capping
layer, a feature that is reproduced by model
fitting.\cite{reflpak} The sample was then cooled to 7 K in a
field below 0.1 mT; and after cooling down the desired field was
applied for measurement. In Figure \ref{reflectivity}(a) we show
the two NSF reflectivity curves together for each field. The polarized reflectivity data measured at 1.5 mT shows that, while the structural Bragg peak was nearly unchanged, a splitting appeared between the two NSF curves at $\approx$ 0.031 \AA$^{-1}$. This provides a signature that there is an additional periodicity with twice
the length of the multilayer period, and it is caused by the spin components parallel or antiparallel to the neutron polarization. It is evident that such magnetic
periodicity is consistent with AFM IEC between the DMS
layers. Note, however, that this splitting is fully suppressed when
the applied field is increased to 100 mT. Instead, a new splitting is observed
at the structural Bragg peak, indicating FM saturation. In
contrast, the undoped sample shows a splitting of only its
structural Bragg peak, even at lowest fields. It indicates that the DMS layers in the undoped sample are aligned ferromagentically along the applied field. We therefore conclude that IEC in the undoped sample is very different from that in the Be-doped, and is either FM or nearly uncoupled.

Using the REFLPAK program,\cite{reflpak} we performed quantitative
fitting of the reflectivity curves. The magnetization in the DMS
layers was assumed to be uniform. The fitting results, plotted as
solid curves in Figure \ref{reflectivity}(a), show that the
reflectivity measured at the two fields is indeed due to
AFM and FM IEC between the ferromagnetic DMS layers,
respectively. The splittings in the $R\times Q^4$ curves are amplified in Figures \ref{reflectivity}(b,c). At
100 mT, which is above the saturation field, we obtain the
magnetic moment per Mn ion projected along the [110] direction to
be $m_\mathrm{Mn}^{[110]}$ = 2.8$\pm$0.3$\mu_\mathrm{B}$. The
difference with respect to the value 3.2$\mu_\mathrm{B}$
previously reported\cite{potashink02} can most probably be
ascribed to the presence of interstitial Mn ions, since our
samples were not annealed to reduce their concentration.\cite{blinowski03} In comparison, we obtain
$m_\mathrm{Mn}^{[110]}$ = 2.1$\pm$0.2$\mu_\mathrm{B}$ for the
AFM-coupled phase at 1.5 mT. The ratio between the observed
moments is consistent with the rotation of the easy axis away from
the [110] direction by $\eta$ = 40$^o \approx$
cos$^{-1}$(2.1/2.8) at 7 K.\cite{welp03,shin07} However, this
rotation does not result in AFM components
oriented along the [1\={1}0] direction, which would have been the case if
we were dealing with the rotation of a single AFM domain. We found no observable
splittings when the sample was oriented with
the [1\={1}0] direction parallel to the neutron polarization and
the same measurement was repeated. We therefore suspect that each DMS layer consists of a distribution of two types of domains whose easy axes are tilted away from [110] by $\pm\eta$, respectively, thus canceling the spin components along [1\={1}0]. Such behavior also suggests that the sample is likely to form a virtually single domain near $T_C$, where uniaxial
anisotropy is dominant.

\begin{figure}
\includegraphics[width=3.2in]{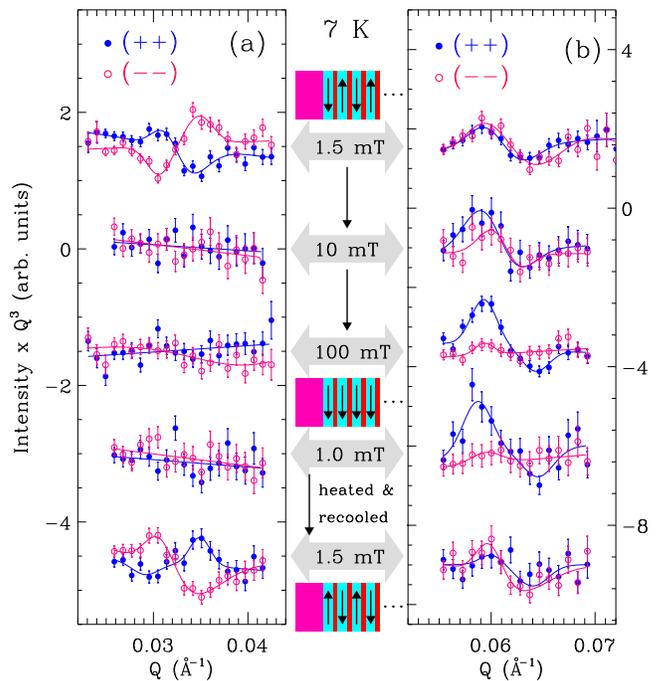}%
\caption{\label{cycle_7k} The AFM(a) and the FM(b) splittings in
the NSF reflection intensities of the Be-doped sample measured at
7 K. The data were collected sequentially in the order from top to
bottom, and corresponding applied fields are shown in the middle. The lines through the data are guides for the eye. The
cartoons in the middle column schematically show the spin
orientations in the multilayer at corresponding fields.}
\end{figure}

\begin{figure}
\includegraphics[width=3.2in]{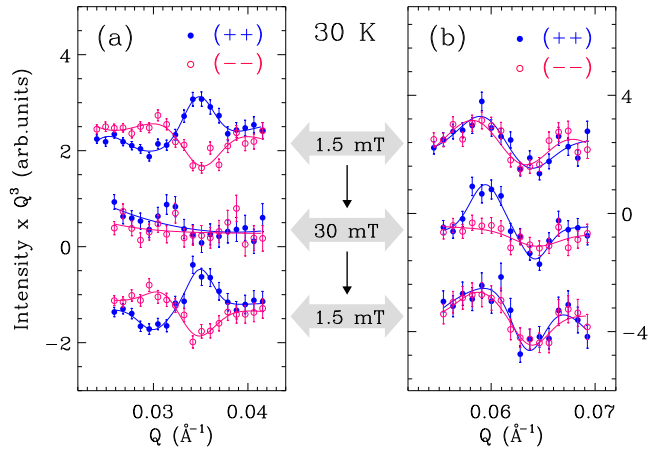}%
\caption{\label{cycle_30k} The AFM(a) and the FM(b) splittings in
the NSF reflection intensities of the Be-doped sample measured at
30 K. The sequence of applied fields is shown in the middle. The lines through the data are guides for the eye.}
\end{figure}

We examined the field dependent behavior of the IEC in more detail
by measuring the NSF reflection intensities. The
plots in Figure \ref{cycle_7k} show that, as the applied field is
increased at 7 K, the FM splitting is enhanced at the expense of
the AFM splitting. The AFM splitting is almost completely suppressed
around 10 mT consistent with the magnetization data, but full development of the FM splitting required
higher fields. When the field is lowered down to 1 mT directly
from full saturation at 100 mT, the AFM splitting does not recover, and only the FM
splitting remains. This is ascribed to a lock-in caused by the
strong cubic biaxial anisotropy field at low
temperatures.\cite{shin07} The AFM IEC can be recovered by raising
the temperature above $T_C$ and then recooling down to 7 K.
We observed that in this process the
direction of the splitting can be reversed, i.e., that the interlayer spin
correlations can change from
$\uparrow\downarrow\uparrow\downarrow\cdot\cdot\cdot$ to
$\downarrow\uparrow\downarrow\uparrow\cdot\cdot\cdot$. This result
indicates that the observed AFM IEC is not initiated by some weak
remanent field during the cooling, but is truly intrinsic to the
sample.

In contrast, when the sample is cooled only to 30 K, the AFM IEC is recovered without a lock-in after field cycling. Figure
\ref{cycle_30k} shows that a nearly identical AFM splitting is observed when the field is raised to 30 mT and lowered back to 1.5
mT. It is because the cubic anisotropy field decreases in strength at higher temperatures, and is not strong enough at 30 K to cause a lock-in of the FM alignmet. All these results show that the AFM IEC observed in our sample is stable over the temperature range observed.

In the case of metal-based multilayers, the IEC between the magnetic layers is known to oscillate between AFM
and FM as a function of the non-magnetic spacer thickness,\cite{parkin90,parkin91} being induced by RKKY-type interaction through
conduction electrons.\cite{bruno91} Since the GaAs is insulating,
carrier injection into the spacers is required to expect similar effects.
The Be doping in the spacers is known to increase the hole
concentrations directly in the GaAs layers\cite{lee04}, as well as
in nearby DMS layers.\cite{wojtowicza03} The hole concentration in
the GaAs:Be layers in our sample is estimated to be 1.0$\times$
10$^{20}$ /cm$^3$ at 7 K. Recent theoretical studies predict
that IEC in GaMnAs-based multilayers can also be changed
from FM to AFM via carrier doping and spacer thickness
control.\cite{sankowski05,giddings06} In light of these works,
however, the discovery of stable AFM IEC in our sample is quite intriguing. The spacers in our sample are as thick as $\approx$ 12
monolayers, a thickness for which according to calculations the exchange strength should be very weak. We suspect that in our case the stability of the observed AFM IEC may have been enhanced by the thickness of the DMS layers, which is $\approx$ 25 monolayers. Interestingly, a mean field calculation shows that, as the thickness of the DMS layer
is increased, the oscillatory behavior of the IEC is changed and
becomes less dependent on the spacer thickness.\cite{giddings06}
Our result therefore may be indicating that reliable magnetoresistance
devices based on DMS multilayers require thicknesses (both for the DMS layers and the spacers) much greater than those typically used for metal-based devices. We expect that continuing studies will establish optimum
design parameters for switchable magnetoresistance devices based on DMS multilayers.


In summary, using DC magnetization and polarized neutron reflectometry measurements, we have observed AFM IEC in Ga$_{0.97}$Mn$_{0.03}$As/GaAs:Be multilayers in which the non-magnetic GaAs spacers are doped with Be. In sharp contrast,
Ga$_{0.97}$Mn$_{0.03}$As/GaAs multilayers with no Be doping showed only FM alignment. Our experimental
finding is thus an important step toward theoretical and
quantitative understanding of IEC between DMS layers separated by
nonmagnetic spacers.

\begin{acknowledgments}
This work was supported by the Korea Science and Engineering Foundation (KOSEF) grant funded by the Korea government (MEST) (No. R01-2008-000-10057-0), and also by the Nuclear R\&D Programs (M20701050003-08N0105-00311). The work at Notre Dame was supported by NSF
Grant DMR06-03752.
\end{acknowledgments}


\end{document}